\documentclass[floatfix,twocolumn,floats,prl,aps,unsortedaddress,superscriptaddress]{revtex4-1}

\usepackage[dvips]{graphicx}
\usepackage{amsfonts}
\usepackage{amsmath}
\usepackage{amssymb}
\usepackage{exscale}
\usepackage{color}	
\usepackage{braket}	
\usepackage{multirow}	
\usepackage{epstopdf}

\usepackage[normalem]{ulem}    

\usepackage[version=3]{mhchem} 
\usepackage[T1]{fontenc}       
\usepackage[utf8]{inputenc}
\usepackage{siunitx}
\DeclareSIUnit\oersted{Oe}
\begin{document}

\title{Two-dimensional electron system at the magnetically tunable EuO/\ce{SrTiO3} interface}

\author{Patrick~L\"omker}
\thanks{P.L. and T.C.R. contributed equally to this work}
\affiliation{Peter Gr\"unberg Institut (PGI-6), Forschungszentrum J\"ulich GmbH, 
			D-52428 J\"ulich, Germany}
\author{Tobias~C.~R\"odel}
\thanks{P.L. and T.C.R. contributed equally to this work}
\affiliation{CSNSM, Univ. Paris-Sud, CNRS/IN2P3, Universit\'e Paris-Saclay, 
			91405 Orsay Cedex, France}
\affiliation{Synchrotron SOLEIL, L'Orme des Merisiers, Saint-Aubin-BP48, 
			91192 Gif-sur-Yvette, France}
\affiliation{Laboratory for Photovoltaics, Physics and Material Science Research Unit, 
			University of Luxembourg, L-4422 Belvaux, Luxembourg}
\author{Timm~Gerber}
\affiliation{Peter Gr\"unberg Institut (PGI-6), Forschungszentrum J\"ulich GmbH, 
			D-52428 J\"ulich, Germany}
\author{Franck~Fortuna}
\affiliation{CSNSM, Univ. Paris-Sud, CNRS/IN2P3, Universit\'e Paris-Saclay, 
			91405 Orsay Cedex, France}
\author{Emmanouil~Frantzeskakis}
\affiliation{CSNSM, Univ. Paris-Sud, CNRS/IN2P3, Universit\'e Paris-Saclay, 			
			91405 Orsay Cedex, France}
\author{Patrick~Le~F\`evre}
\affiliation{Synchrotron SOLEIL, L'Orme des Merisiers, Saint-Aubin-BP48, 
			91192 Gif-sur-Yvette, France}
\author{Fran\c{c}ois~Bertran}
\affiliation{Synchrotron SOLEIL, L'Orme des Merisiers, Saint-Aubin-BP48, 
			91192 Gif-sur-Yvette, France}
\author{Martina~M\"uller}
\email{mart.mueller@fz-juelich.de}
\affiliation{Peter Gr\"unberg Institut (PGI-6), Forschungszentrum J\"ulich GmbH, 
			D-52428 J\"ulich, Germany}
\affiliation{Fakult\"at Physik, Technische Universit\"at Dortmund, D-44221 Dortmund, Germany}
\author{Andr\'es~F.~Santander-Syro}
\email{andres.santander@csnsm.in2p3.fr}
\affiliation{CSNSM, Univ. Paris-Sud, CNRS/IN2P3, Universit\'e Paris-Saclay, 
			91405 Orsay Cedex, France}

\date{\today}

\begin{abstract}
	We create a two-dimensional electron system (2DES) 
	at the interface between EuO, a ferromagnetic insulator, 
	and SrTiO$_3$, a transparent non-magnetic insulator considered the bedrock 
	of oxide-based electronics. 
	This is achieved by a controlled \emph{in situ} redox reaction 
	between pure metallic Eu deposited at room temperature on the surface of SrTiO$_3$
	-- an innovative bottom-up approach that can be easily generalized 
	to other functional oxides and scaled to applications.
	Additionally, we find that the resulting EuO capping layer 
	can be tuned from paramagnetic to ferromagnetic, 
	depending on the layer thickness.
	These results demonstrate that the simple, novel technique of creating 2DESs
	in oxides by deposition of elementary reducing agents
	[T.~C.~R\"odel \emph{et al.}, Adv. Mater. \textbf{28}, 1976 (2016)]
	can be extended to simultaneously produce an \emph{active}, e.g. magnetic,
	capping layer enabling the realization and control of additional functionalities 
	in such oxide-based 2DESs.
\end{abstract}
\maketitle

{\it Introduction.-} 
Two-dimensional electron systems (2DESs) in functional oxides 
have gained strong interest 
as a novel state of matter with fascinating and exotic interface physics. 
For instance, the 2DES in LaAlO$_3$/SrTiO$_3$ (LAO/STO) interfaces can host 
metal-to-insulator transitions, superconductivity and magnetism --all of them tunable 
by gate electric fields~\cite{Ohtomo2004,Thiel2006,Ueno2008,Nakamura2009, 
Reyren2007,Brinkman2007,Caviglia2008,Caviglia2010,Joshua2013}.
The prospect of creating and manipulating a macroscopic magnetic ground state 
in oxide-based 2DESs is of enormous interest, as this would pave the route towards 
oxide spintronic applications with novel quantum phases 
beyond today's semiconductor technology.

Recent studies aimed at supporting the existence of magnetic ordering 
at the LAO/STO interface, e.g. by the observation 
of tunnel magnetoresistance (TMR)~\cite{Ngo2015} or the 
inverse Edelstein effect~\cite{Lesne2016,Song2016}. 
The magnetic field dependence of TMR was attributed to a Rashba-type 
spin-orbit coupling, potentially allowing the manipulation of spin polarization 
in a 2DES, whereas its spin-momentum locking may enable a high efficiency 
of the conversion of an injected spin current into a charge current.  
In fact, in the case of the LAO/STO interface, it was recently demonstrated
that additional epitaxial ferroic oxide layers can be used to tune 
the spin polarization of the 2DES by an electric field~\cite{Stornaiuolo2015} 
or to control its conduction in a non-volatile manner 
by ferroelectric switching~\cite{Tra2013}.

So far, the design of functional 2DES required a single layer growth control 
of epitaxial LAO onto SrTiO$_3$. 
The emergence of interfacial quantum states, such as magnetism,  
superconductivity or spin-orbit coupling, only sets in at a critical LAO thicknesses 
of $4$ unit cells and in certain regions of  the 2DES phase diagram~\cite{Ngo2015}.
This conundrum was circumvented by the finding that 2DESs could be fabricated 
at the bare surface of several oxides, through the creation of oxygen vacancies 
at their surface~\cite{Santander-Syro2011,Meevasana2011,Santander-Syro2012,Bareille2014,
Santander-Syro2014,Roedel2014,Roedel2015,Frantzeskakis2016}. 
These surface 2DESs can also show magnetic domains~\cite{Taniuchi2016},
thus constituting an appealing alternative for the use and control
of electric and magnetic properties of confined states in oxides.

\begin{figure}[!tb]
	\begin{center}
		\includegraphics[clip, width=0.48\textwidth]{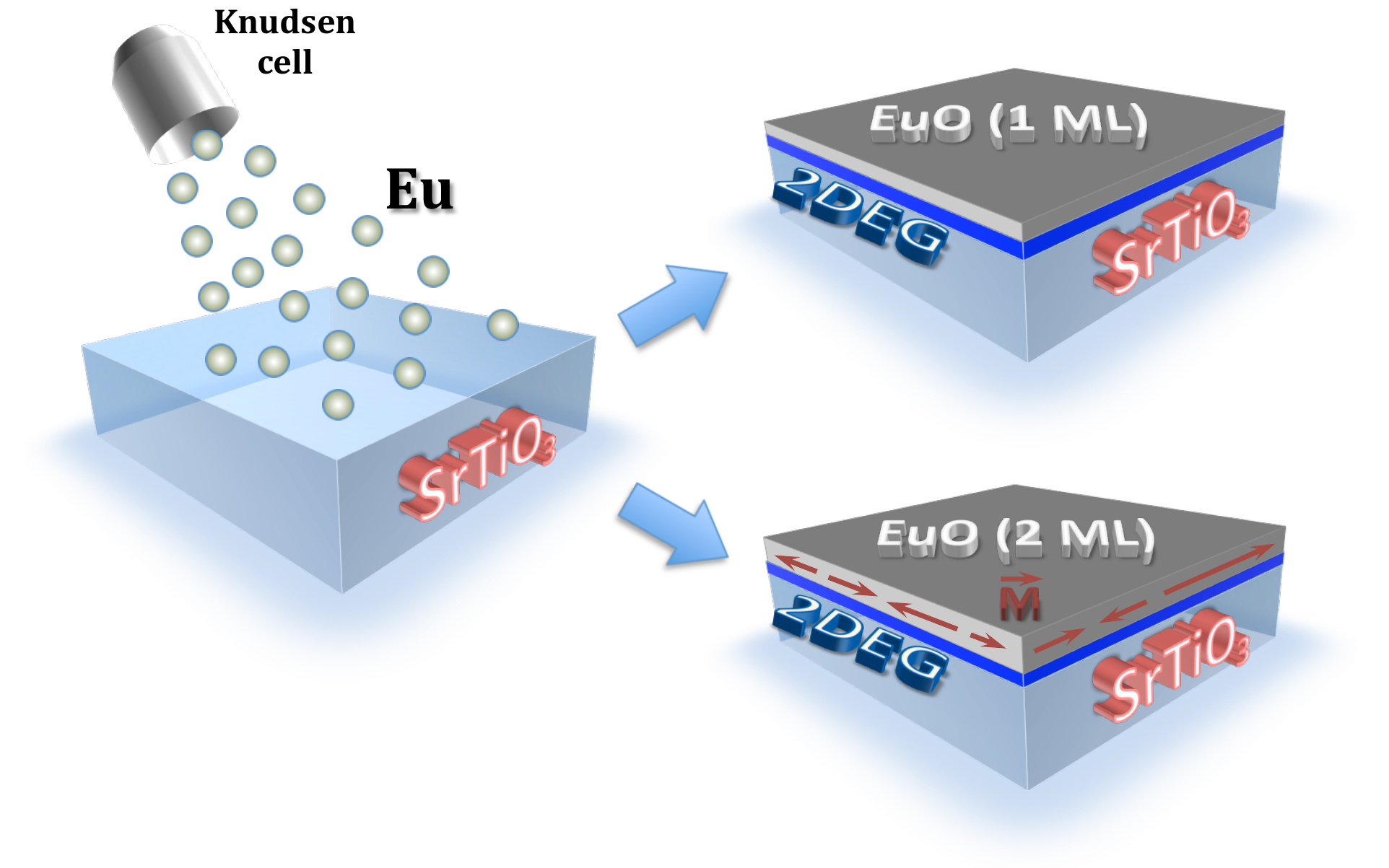}
	\end{center}
	\caption{\footnotesize{
			Schematics of the experiment. Pure Eu metal (grey balls), 
			evaporated from a Knudsen cell, reacts with the \ce{SrTiO3} surface, 
			forming stoichiometric insulating EuO (grey). The redox reaction locally
			reduces the \ce{SrTiO3} around its surface, creating a 2DEG (blue).
			The capping layer can be tuned from paramagnetic, for 1~ML of EuO,
			to ferromagnetic, for 2~ML of EuO -- where magnetic domains
			with in-plane magnetization 
			$\mathbf{M}$ are represented by the red arrows.
			}
	}
\label{fig:fig1}
\end{figure}

Here, we show that insulating and ferromagnetic EuO can be grown on SrTiO$_3$ 
while simultaneously creating a 2DES at the interface. 
As schematized in Fig.~\ref{fig:fig1},
the fabrication of the 2DES is simply accomplished by the deposition of 
pure metallic Eu at room temperature in ultrahigh vacuum. 
We find that the resulting EuO capping layer can be tuned 
from paramagnetic to ferromagnetic, depending on the Eu metal coverage 
($d_{Eu} = 1$~ML and 2~ML, respectively), and show, 
using angle-resolved photoemission spectroscopy (ARPES),
that the integrity of the 2DES is preserved in both cases,
thus providing an ideal knob for tuning the spin-transport properties
of the 2DES. 
This bottom-up approach to create a 2DES by an interfacial redox process 
relies on recent results demonstrating that the evaporation 
of an amorphous ultra-thin layer of Al metal on top of an oxide surface 
generates a homogeneous 2DES~\cite{Roedel2016}. 
As the redox reaction between oxides and elementary metals 
with a large heat of formation of the corresponding metal oxide 
is a general phenomenon~\cite{Campbell1996}, 2DESs can be created in various oxides, 
e.g. SrTiO$_3$, TiO$_2$, and BaTiO$_3$~\cite{Roedel2016}.  
In the present study, we advance this exciting possibility 
towards simultaneously creating a 2DES and forming a \emph{functional} 
metal oxide overlayer --i.e. in a macroscopic ferromagnetic ground state--
by choosing a suitable elementary metal (Eu). 
Our experiments demonstrate how to elegantly link the simplicity and universality 
of an interfacial redox reaction to obtain increased functionalities 
by engineering \emph{just one} active oxide overlayer that can enhance, modify 
and allow controlling the properties of the subjacent so-created confined electron system. 

{\it Methods.-}
The preparation of ultrathin EuO films by oxide 
molecular beam epitaxy (MBE) poses several experimental 
challenges~\cite{Mueller2009,Caspers2013,Gerber2016,Steeneken2002,Ulbricht2008,
Sutarto2009,Foerster2001,Caspers2011, Caspers2016, Prinz2016}.
The oxygen partial pressure, the substrate temperature, 
and the rate of impinging Eu-metal atoms must be carefully controlled. 
However, only the stoichiometric compound yields the desired 
simultaneous occurrence of magnetic and semiconducting behaviors. 
In this paper a novel method to synthesize ultrathin EuO
is demonstrated and put into practice, \emph{i.e.} a controlled interfacial redox reaction 
with oxygen provided by the substrate material \emph{only}~\cite{Loemker2017}. 

The undoped TiO$_2$-terminated \ce{SrTiO3} samples are prepared 
using a well established technique~\cite{Koster1998}.
Atomic force microscopy images show 
a flat surface with steps of unit cell height and a roughness 
within one terrace of typically \SI{150}{\pico\meter} 
and a $c$-direction miscut angle $<0.1^{\circ}$. 
The samples are then annealed in vacuum to \SI{500}{\celsius} for \SI{0.5}{\hour} 
in a MBE chamber at a base pressure of \SI{1.3e-10}{\milli\bar}
prior to Eu evaporation and photoemission experiments.
The cleanliness and crystallinity of the so-obtained surfaces are checked by
\emph{in situ} X-ray photoemission spectroscopy (XPS). 
Pure Eu metal is then evaporated at \SI{480}{\celsius} 
at a rate of \SI{0.3}{\angstrom\per\minute} 
using a low temperature Knudsen cell, while the \ce{SrTiO3} substrate 
is kept at room temperature. 
The deposition rate of Eu metal is monitored by a calibrated quartz microbalance. 

The redox-created oxidation state of the Eu on the \ce{SrTiO3} surface 
is analyzed using XPS with Al~K$_{\alpha}$ radiation from a SPECS X-Ray anode 
and a PHOIBOS-100 hemispherical energy analyzer at FZ J\"ulich. 
The Eu $3d$ and Ti $2p$ core-levels are analyzed to quantify the oxidation state 
of the deposited Eu-metal and to observe the redox process with the substrate surface.
Before the \textit{ex situ} magnetization measurements, 
realized with a Quantum Design MPMS SQUID magnetometer,
the EuO/SrTiO$_3$ samples are further capped with \SI{15}{\nm} 
of e-beam evaporated MgO to avoid additional oxidation. 
A hysteresis loop of $H=\pm\SI{1500}{\oersted}$ at $T=\SI{5}{\K}$ is performed, 
while temperature dependence is recorded with an aligning field 
of $H = \SI{500}{\oersted}$ for $T=$\SIrange{5}{150}{\K}.
All magnetization data was measured in-plane. 

The ARPES measurements are conducted at the CASSIOPEE beamline of synchrotron SOLEIL. 
The beamline is equipped with an MBE chamber 
allowing the \textit{in situ} preparation of the \ce{SrTiO3} surfaces 
and evaporation of the pure Eu-metal using the same above-specified conditions. 
We furthermore checked that a surface cleaning using a much faster annealing 
(about one minute) creates a negligible amount of bulk oxygen vacancies.
Eu evaporated hereafter then results in an identical 2DES,
in line with previous reports showing that the electronic structure 
of the 2DES at the surface of SrTiO$_3$ is independent 
of the material's bulk doping~\cite{Santander-Syro2011}.
We used linearly polarized photons at energies of $47$~eV and $90$~eV, 
which provide the best cross-section for ARPES spectra on 
\ce{SrTiO3}~\cite{Santander-Syro2011, Roedel2016},
and a hemispherical electron analyzer with vertical slits.
The angular and energy resolutions were $0.1^{\circ}$ and 8~meV. 
The mean diameter of the incident photon beam was smaller than \SI{100}{\micro\meter}.
The samples were measured at $T=8$~K.
The results were reproduced on two samples.
All through this paper, we note $\langle hkl \rangle$ 
the directions in reciprocal space.
The indices $h$, $k$, and $l$ correspond to the reciprocal lattice vectors 
of the cubic unit cell of SrTiO$_3$.

\begin{figure}[!tb]
	\begin{center}
		\includegraphics[clip, width=0.48\textwidth]{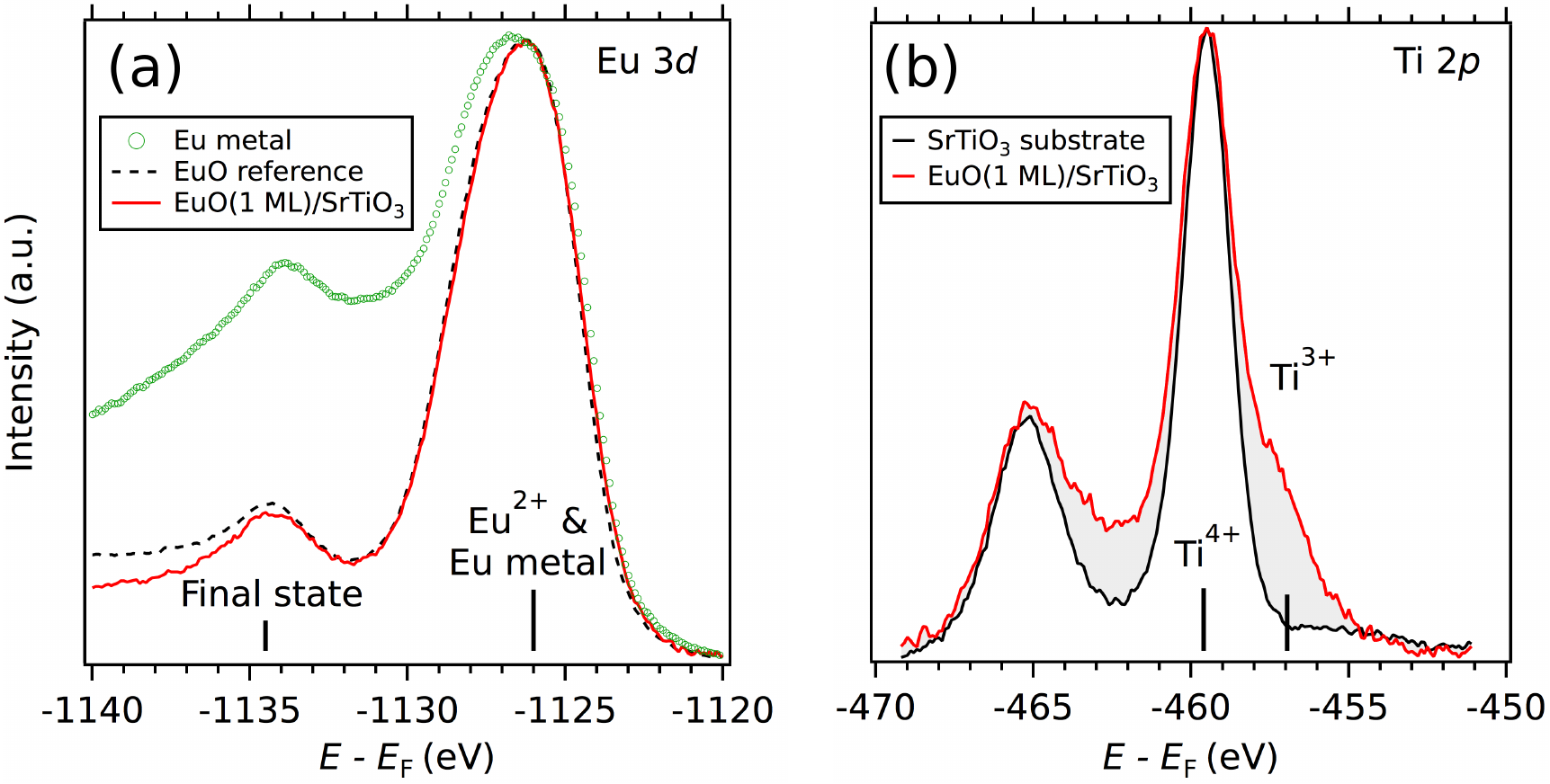}
	\end{center}
	\caption{\footnotesize{
			XPS data of (a) the Eu~$3d_{5/2}$ peak and 
			(b) the Ti $2p$  core level. 
			Both (a) and (b) illustrate the interfacial redox process
			after evaporation of pure Eu metal on the surface of SrTiO$_3$: 
			oxygen provided from the substrate forms 1ML of stoichiometric EuO, 
			while the Ti of the substrate is reduced to Ti$^{3+}$. 
			}
	}
\label{fig:fig2}
\end{figure}
{\it Results.-}
To show that ultrathin stoichiometric EuO can be grown 
without supplying additional oxygen, the films are analyzed using XPS. 
Fig.~\ref{fig:fig2}(a) shows the Eu~$3d_{5/2}$ core-level. 
The chemistry of the film can be determined by comparison to reference spectra 
of Eu-metal, Eu$^{2+}$, and Eu$^{3+}$. The dashed line represents stoichiometric EuO. 
In accordance with previous studies of the Eu~$3d$ core level, the Eu$^{2+}$ valence 
is located at an energy of $-\SI{1125}{\eV}$~\cite{Caspers2011, Gerber2016, Cho1995}. 
The peak is accompanied by a well known satellite at higher binding energy, 
which is part of the multiplet of the $3d^9$~$4f^7$ final state~\cite{Cho1995}.
The red line shows the spectrum of a SrTiO$_3$ sample with \SI{4}{\angstrom} of Eu-metal 
deposited on top of it. The amount of Eu metal corresponds to a thickness 
of $\approx 1$~ML of EuO.
We find that the XPS spectra of our Eu-capped SrTiO$_3$ samples 
is indistinguishable from stoichiometric EuO reference data. 
Features related to Eu-metal or Eu$^{3+}$ are absent.
Analogous results are found (not shown) in case of a deposited Eu-metal layer 
of \SI{8}{\angstrom}.
This demonstrates that, in the ultrahigh vacuum conditions used here, 
the Eu metal is oxidized into EuO at the surface of SrTiO$_3$.

The concomitant substrate reduction is evidenced by the analysis
of the Ti~$2p$ core level, shown in Fig.~\ref{fig:fig2}(b).
For stoichiometric \ce{SrTiO3} a pure Ti$^{4+}$ valence is observed. 
Upon deposition of nominally 1~ML of EuO, the XPS spectrum shows an additional component 
at the binding energy of Ti$^{3+}$, indicating that the \ce{SrTiO3} is indeed reduced. 


\begin{figure}[!tb]
	\begin{center}
		\includegraphics[clip, width=0.4\textwidth]{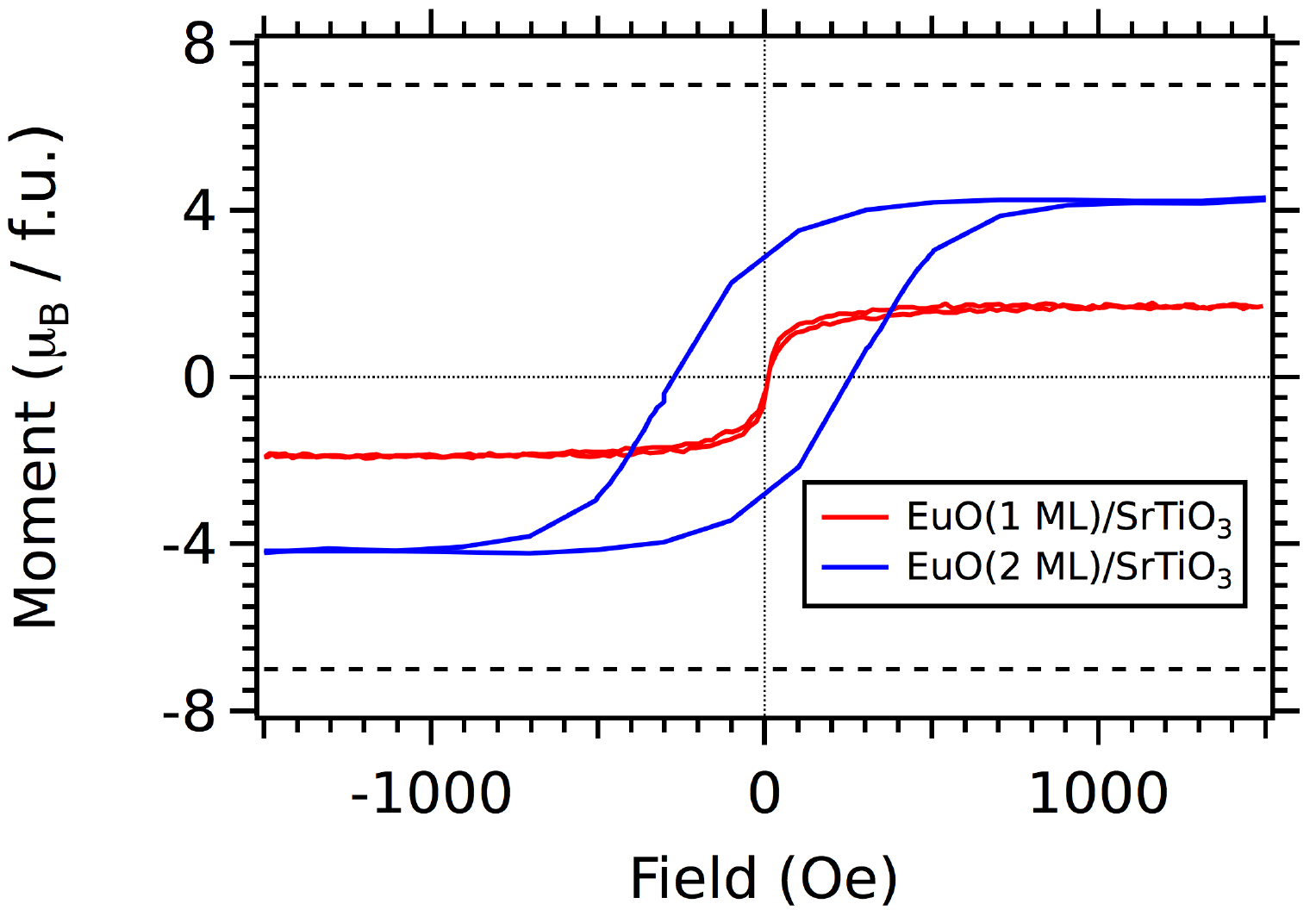}
	\end{center}
	\caption{\footnotesize{ 
			In-plane magnetization curves measured at $T=5$~K,
			obtained with a Quantum Design MPMS SQUID magnetometer,
			contrasting the paramagnetic behavior of 1~ML EuO 
			and the low-dimensional ferromagnetic behavior of 2~ML EuO 
			formed after evaporating pure Eu metal
			on TiO$_2$ terminated SrTiO$_3$.
			The horizontal dashed lines show the expected saturation magnetization 
			at $T=0$~K for EuO~\cite{Schiller2001,Mueller2009}.
			}
	}
\label{fig:fig4}
\end{figure}
The unique properties of the obtained capping EuO layer, 
and their tuning with layer thickness, are presented in Fig.\,\ref{fig:fig4}. 
The measured (not shown) ferromagnetic transition temperature of the $2$~ML EuO film 
was $T \approx 60$~K.
At $T=5$~K, the magnetization versus field $M(H)$ curve of 1~ML of EuO (red curve)
shows a paramagnetic behavior. 
In this case, the effective coordination number of Eu atoms is lower 
compared to the coordination number in bulk EuO, 
and thus exchange interactions are weakened~\cite{Mueller2009}. 
However, at the same temperature, 
the 2~ML EuO overlayer is ferromagnetic (blue curve) 
with a saturation magnetization of $M_S = 4 \mu_{B}$ per formula unit (f.u.).
The measured saturated magnetic moment for $2$~ML of EuO capping is close 
to the corresponding theoretical values for EuO at $T=0$~K,
represented by the horizontal dashed 
lines~\cite{Schiller2001,Mueller2009,Caspers2013,Gerber2016}.
Now the underlying 2DES (see next) is interfaced with a magnetic material,
which may ultimately enable a control of the spin degrees of freedom in this system. 

\begin{figure}[!tb]
	\begin{center}
		\includegraphics[clip, width=0.46\textwidth]{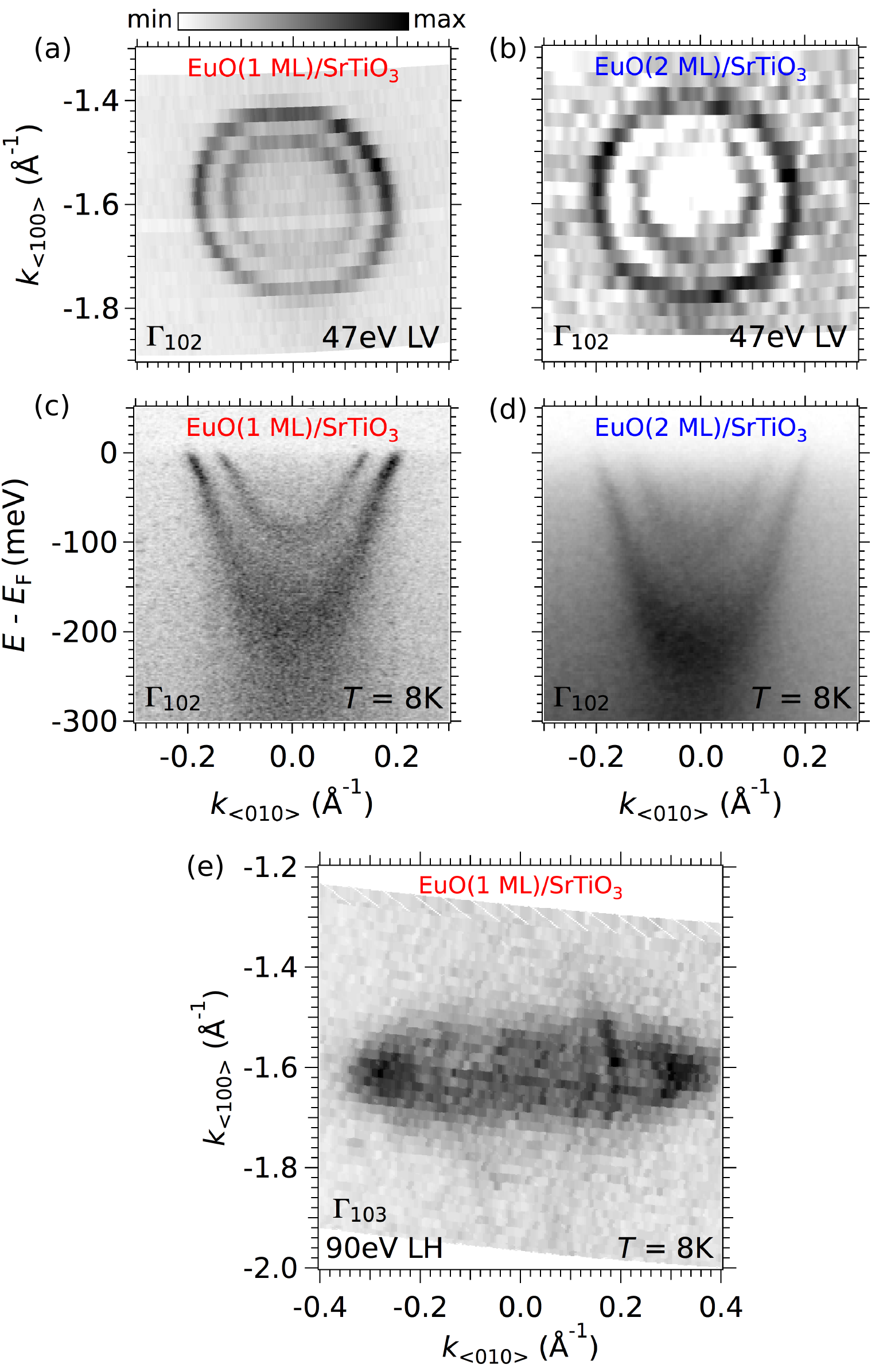}
	\end{center}
	\caption{\footnotesize{
			ARPES data of the EuO/\ce{SrTiO3} interface. 
			(a,~b) Fermi-surfaces taken around the $\Gamma_{102}$ point 
			of SrTiO$_3$ for nominally 1~ML (raw data) and 2~ML 
			(negative values of second derivatives) EuO coverage, respectively,
			using $47$~eV photons with linear vertical (LV) light polarization. 
			These photon energy and polarization enhance the photoemission intensity
			of the $3d_{xy}$ circular Fermi surfaces.
			(c~d) Corresponding dispersion of the two Ti $3d_{xy}$ light subbands.
			(e) Fermi-surfaces taken around the $\Gamma_{103}$ point 
			of SrTiO$_3$ for 1~ML EuO coverage using $90$~eV photons 
			with linear horizontal (LH) light polarization.
			These photon energy and polarization enhance the photoemission intensity
			of one of the two orthogonal $3d_{xz/yz}$ ellipsoidal Fermi surfaces.
			As with our previous results on Al-capped SrTiO$_3$~\cite{Roedel2016}, 
			we crosschecked that for both 1~ML and 2~ML EuO 
			the 2DEG forms instantaneously after the Eu deposition,
			and its carrier density is independent of the dose of UV light 
			used to measure the ARPES data. 
			In other words, the 2DEG is entirely due to the oxidation of the capping layer.
			}
	}
\label{fig:fig3}
\end{figure}
The formation of oxygen vacancies near the SrTiO$_3$ surface, 
induced by the redox reaction with the Eu evaporated on top of it, 
results in a local electron doping of the substrate and the creation of a 2DES,
in analogy with the 2DESs formed by oxygen vacancies at the UV-irradiated surface
or Al-capped interface of SrTiO$_3$ or other oxides~\cite{Santander-Syro2011,Meevasana2011,
Santander-Syro2012,Bareille2014,Roedel2015,Roedel2016}.
This is directly demonstrated by the ARPES data shown in Fig.~\ref{fig:fig3}. 
Figs.~\ref{fig:fig3}(a,~b) show the circular Fermi surfaces around $\Gamma_{102}$
of the two $3d_{xy}$ subbands at the interfaces between 1~ML and 2~ML EuO films 
on \ce{SrTiO3}, respectively.
Figs.~\ref{fig:fig3}(c,~d) present the corresponding energy-momentum 
ARPES intensity maps along the $k_{<010>}$ direction at $k_{<100>}=2\pi/a$
($a=3.905$~\AA~is the lattice parameter of SrTiO$_3$). 
These correspond to the two Ti $3d_{xy}$ light subbands previously reported 
for the 2DES in SrTiO$_3$~\cite{Santander-Syro2011,Meevasana2011,Roedel2016}.
For the 2DES at the EuO(1~ML)/SrTiO$_3$ interface,
the band bottoms ($E_0 \approx -200$~meV and $-90$~meV 
for the outer and inner subbands, respectively), 
Fermi momenta ($k_F \approx 0.19$~\AA$^{-1}$ and $0.12$~\AA$^{-1}$),
effective masses ($m^{\star}/m_e = 0.7 \pm 0.05$ for both subbands,
estimated from a parabolic approximation to the band dispersions,
where $m_e$ is the bare electron mass), 
and the observation of a kink at $E \approx -30$~meV 
below $E_F$, ascribed to a band renormalization due to electron-phonon interaction,
are all in agreement with previous reports~\cite{Santander-Syro2011,Meevasana2011,
Roedel2016,Chen2015,Wang2016}.
As shown in Fig.~\ref{fig:fig3}(e), the ellipsoidal Fermi surfaces,
associated with the Ti $3d_{xz/yz}$ heavy subbands, 
are also observed using horizontal light polarization. 
From the total area $A_F$ enclosed by \emph{all} the Fermi surfaces, 
the density of carriers of the 2DES at the EuO/SrTiO$_3$ interfaces is 
$n_{2D} = A_F/(2\pi^2) \approx 2.0 \times 10^{14}$~cm$^{-2}$,
which is comparable to the density of states of the 2DES 
at the bare SrTiO$_3$ surface~\cite{Santander-Syro2011,Roedel2016}.
The thickness of the 2DES can be directly inferred from the number of subbands, 
their band bottoms and energy separations~\cite{Santander-Syro2011}. 
Thus, as the electronic structure of the 2DES at the EuO(1~ML)/SrTiO$_3$ interface 
is essentially the same as the one observed at the bare surface 
of SrTiO$_3$~\cite{Santander-Syro2011}, or at the Al-capped surface of 
SrTiO$_3$~\cite{Roedel2016}, we conclude that its thickness is also the same, 
namely about $4-5$ unit cells.

ARPES measurements are performed under zero external magnetic field,
to guarantee conservation of the photo-emitted electron momentum.
Thus, while magnetizing the capping EuO film is not feasible for these
measurements, it is nevertheless instructive to compare the ARPES data 
between the 1~ML and 2~ML EuO films, Figs.~\ref{fig:fig3}(a,~c) and (b,~d),
respectively. 
Note that, while the Fermi momenta of the 2DESs in both systems are essentially identical,
one observes a small but distinct difference in their bandwidths and band splittings.
Specifically, in the case of the 2~ML EuO film, the bottoms of the Ti $3d_{xy}$ 
light subbands are at about $-230$~meV and $-100$~meV, with a concomitant band splitting
($\approx 130$~meV) slightly larger than the one in the 1~ML EuO film ($\approx 100$~meV).
Additionally, both the Fermi edge at $E=0$ 
and the kink at $E \approx -30$~meV appear much less pronounced in the 
2DES at the EuO(2~ML)/SrTiO$_3$ interface.
The possible link between such differences in electronic structure, and the 
ferromagnetism (with or without domains) in the 2~ML EuO film, 
should be further explored in future works.
On the other hand, an important conclusion at this point is that the onset of 
ferromagnetism in the zero-field-cooled 2~ML EuO films, 
with the concomitant formation of randomly oriented ferromagnetic domains 
(as schematized in Fig.~\ref{fig:fig1}), still preserves the integrity 
of the underlying 2DES. Together with the magnetization data from
Fig.~\ref{fig:fig4}, our results open the very exciting perspective of 
enabling the continuous tuning, under external applied field, 
of the spin transport properties in oxide-based 2DES. 

{\it Conclusions.-}
In summary, we demonstrated that the deposition in vacuum, at room temperature,
of Eu-metal on \ce{SrTiO3} results in the simultaneous creation of a
2DES in the oxide substrate and a capping EuO layer that can be tuned from
paramagnetic (1~ML thickness) to ferromagnetic (2~ML). 
These results open new perspectives for investigating the interaction 
of the magnetic and electronic properties of the 2DES in SrTiO$_3$.
More generally, these results lay a new ground for the simple and versatile
design of all-oxide devices in which the functionalities of the constituting elements, 
and their mutual coupling, can be obtained from controlled physicochemical reactions 
and vacancy engineering at their interfaces.

\acknowledgments 
We thank O. Petracic and the J\"ulich Centre for Neutron science 
for providing measurement time at the SQUID magnetometer.
Work at CSNSM was supported by public grants 
from the French National Research Agency (ANR), 
project LACUNES No ANR-13-BS04-0006-01, 
and the ``Laboratoire d'Excellence Physique Atomes Lumi\`ere Mati\`ere'' 
(LabEx PALM projects ELECTROX and 2DEG2USE) overseen by the ANR as part of the 
``Investissements d'Avenir'' program (reference: ANR-10-LABX-0039).
M.~M. acknowledges financial support from HGF under contract No. VH-NG-811.
T.~C.~R. acknowledges funding from the RTRA--Triangle de la Physique (project PEGASOS).
A.F.S.-S. thanks support from the Institut Universitaire de France.


\end{document}